\documentclass{moriond}

\def\Journal#1#2#3#4{{#1} {\bf #2}, #3 (#4)}

\def\aap{\em A$\&$A}
\def\JCP{\em J. Co. Ph. }

\def\be{\begin{equation}}
\def\ee{\end{equation}}
\def\bea{\begin{eqnarray}}
\def\eea{\end{eqnarray}}

\newcommand{\Photo}{}

\begin{document}
\vspace*{4cm}
\title{Can dark energy explain the observed outflow in galaxy clusters?}

\author{ M. Donnari$^{1,2}$, M. Arca-Sedda$^{1}$, M. Merafina$^{1}$}

\address{$^{1}$Department of Physics, Sapienza University of Rome\\
Piazzale Aldo Moro 5, I-00185 Rome, Italy}

\address{$^{2}$Department of Physics, Tor Vergata University of Rome\\
Via O. Raimondo 18, I-00173 Rome, Italy}

\maketitle\abstracts{
Recent observations of the Virgo cluster and the Local Group suggested that some galaxies are flowing out from their parent cluster. This may be the signature that dark energy (DE) acts significantly also on small cosmological scales. By means of direct N-body simulations we performed several simulations, in which the effect of DE and gravity are taken into account, aiming to determine whether DE can produce an outflow of galaxies compatible with observations. Comparing the different simulations, our results suggest that the observed outflow of galaxies is likely due to the local effect of DE.}

\section{Scientific context and numerical method}
Recent observations of the local velocity field of the Local Group and the Virgo cluster, have revealed a linear
velocity-distance relation of the outermost galaxies, properly referred to as Local Hubble
Flow (LHF) \cite{2002Kara,2010chernin}. 
Following a non-Friedmann cosmology \cite{2006chernin}, we performed a series of direct N-body simulations of a galaxy cluster (composed by 240 galaxies) which suffers the anti-gravitational effect of DE, using a modified version of the direct N-body code \texttt{HiGPUs} \cite{2013spera}. We modelled also the
Intra-Cluster Medium (ICM), limiting the study to its gravitational action, thus avoiding other effects.
Nowdays there are still controversial opinions about the nature of the LHF.
On large scales the Universe is homogeneous and the global Hubble constant is $H_0 \simeq$ 72 km/s/Mpc. On the other hand, on small scales the Universe is inhomogeneous due to
the presence of galaxies and clusters. Because of this, it is possible to define a local
Hubble constant $H_L$, slightly different from the global one, $H_0$.

Each galaxies has been modelled according to the so-called $\gamma$-model and we put them in the space according to a King density profile.
For DE, we take into account the $\Lambda$CDM model, by considering $\rho_{\Lambda}=0.7\times 10^{-29} \; \rm g\; cm^{-3}$ \cite{2014Planck} whereas for ICM we assume that its distribution is well represented by a modified $\beta$-model.
In order to make our results as reliable as possible, we
performed a single-particle (SP) simulation, in which each
galaxy is represented as a point-like mass. This choice allowed us to simulate the evolution of the cluster up to $\simeq$ 30 Gyr, since the computational time has been considerably reduced.

\section{Results}
We studied the trajectory of the outermost galaxies of the cluster founding that after a Hubble time the system lose around 10 $\%$ of its initial mass.
Comparing MM4 and MM1 models (see left panel of \ref{tab:1}), it is clear that the action of DE overtakes
gravity, pushing the galaxy away from the cluster. Even adding the
gravitational effect of ICM, the galaxy still departs from its parent
cluster, confirming that the gas gravitational field poorly affects the
dynamics of the cluster.
The correlation between the radial velocity and the distance from the cluster centre is a way to highlight the effect of
DE on local volume by rougly estimate the value of $H_L$.
\begin{figure}
\begin{minipage}{0.33\linewidth}
\centerline{\includegraphics[width=\linewidth]{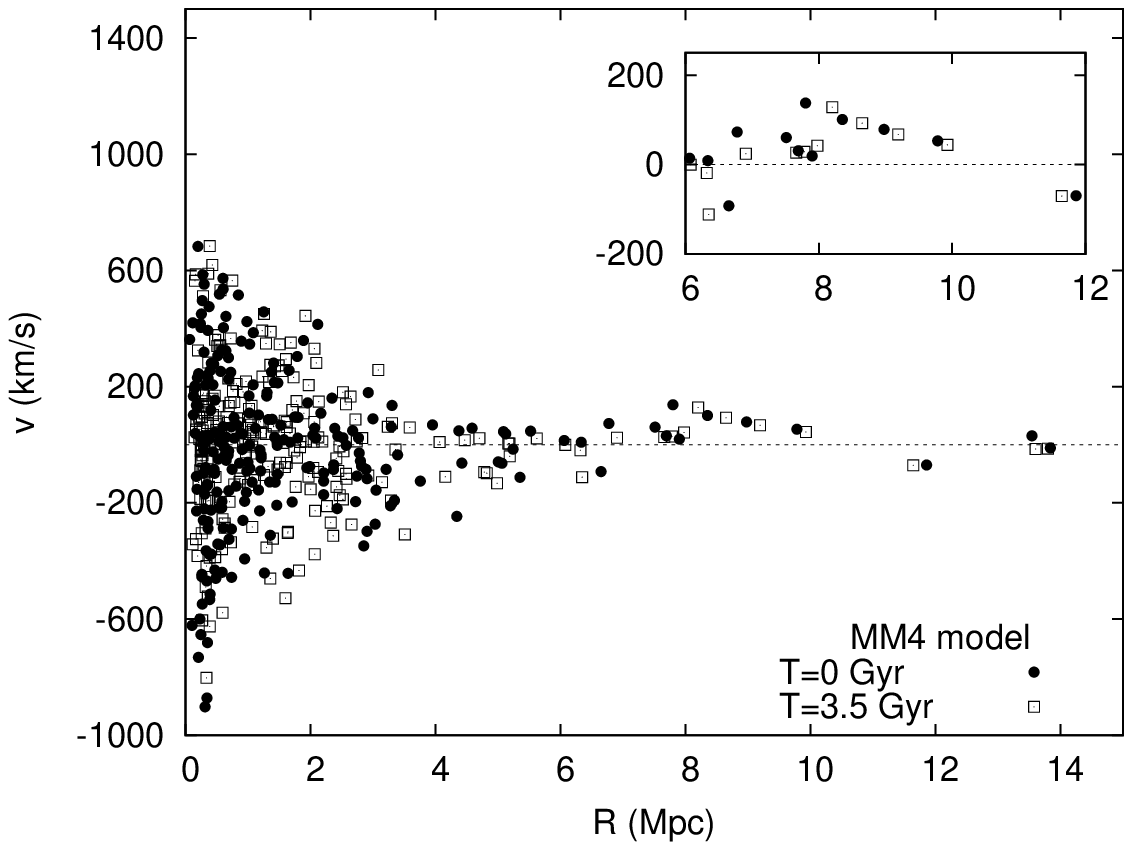}}
\end{minipage}
\hfill
\begin{minipage}{0.32\linewidth}
\centerline{\includegraphics[width=\linewidth]{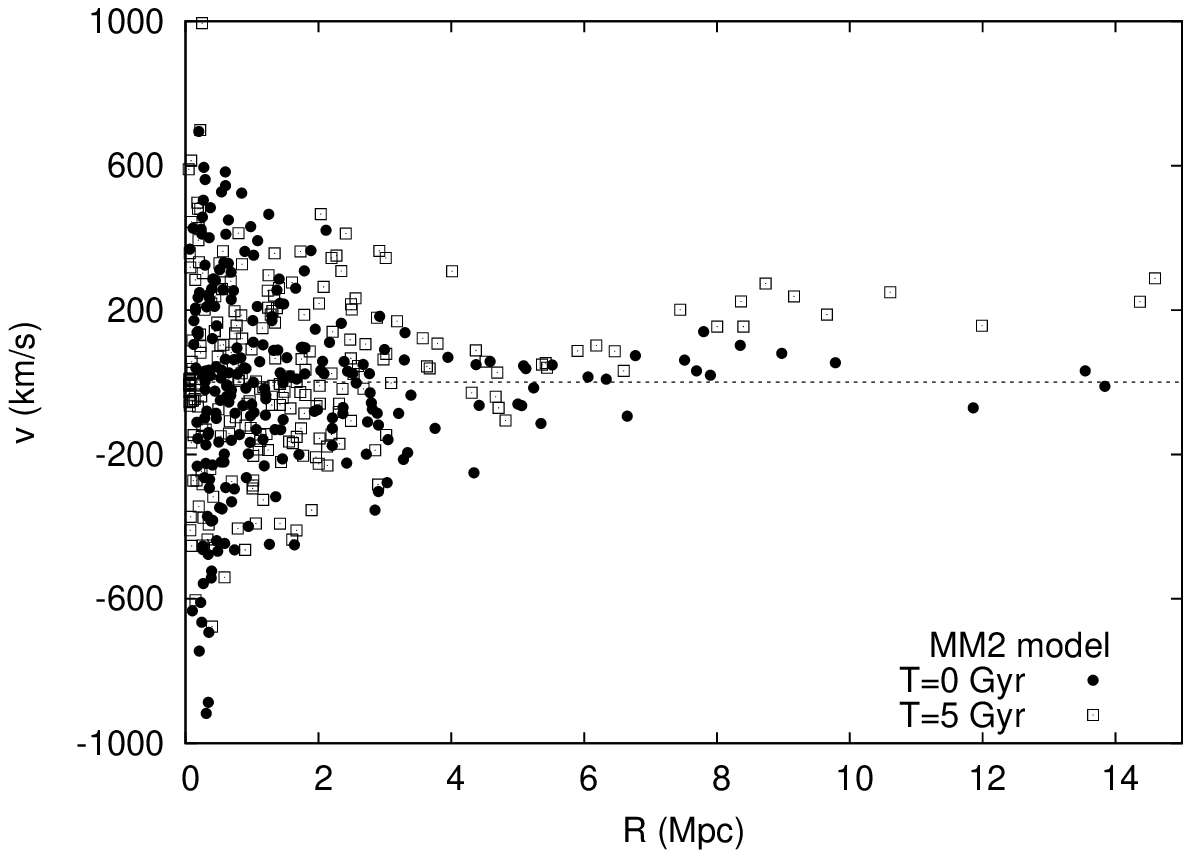}}
\end{minipage}
\hfill
\begin{minipage}{0.32\linewidth}
\centerline{\includegraphics[width=\linewidth]{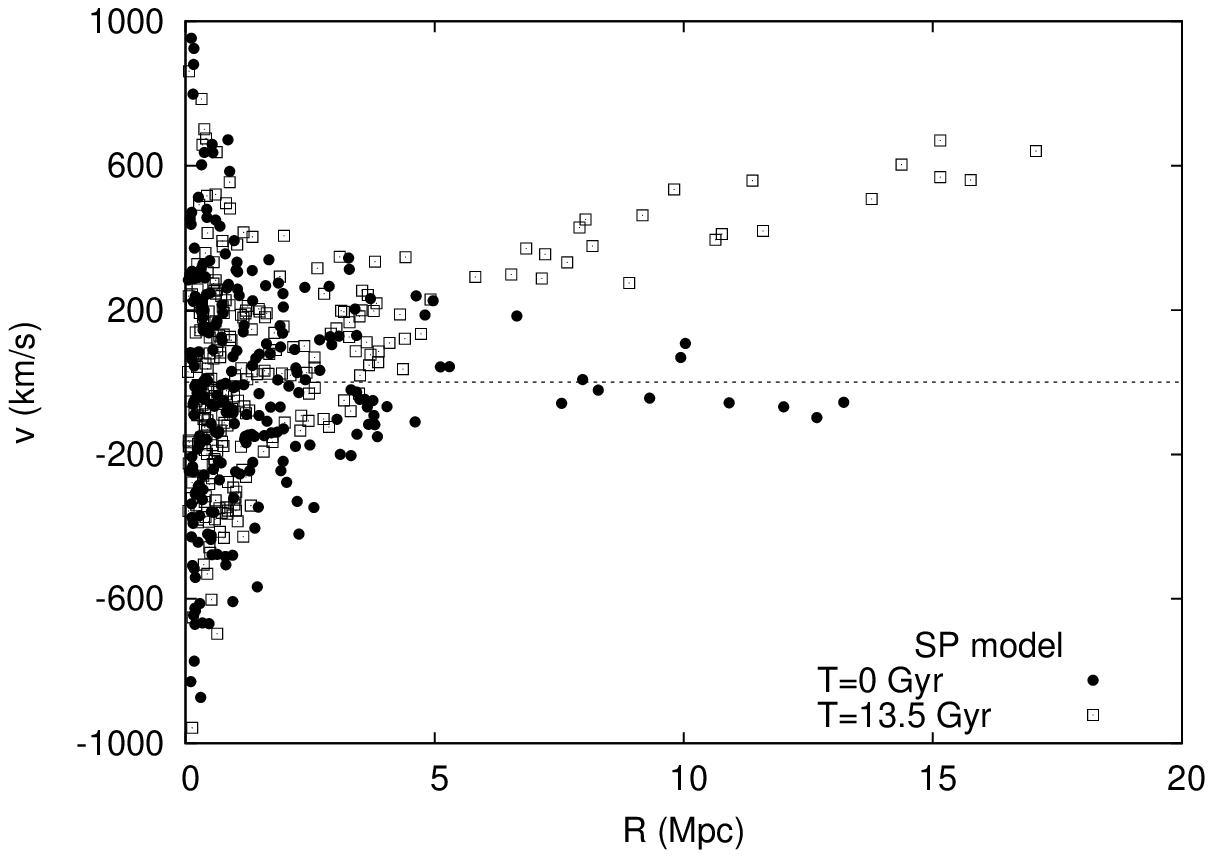}}
\end{minipage}
\caption{Hubble diagrams of the cluster in MM4, MM2 and SP models. Dots represent the starting point wherease open squares represent the end of the simulation.}
\label{fig:Hubble-diagrams}
\end{figure}
In Fig. \ref{fig:Hubble-diagrams} is shown the Hubble diagram for MM4 model (left) in which the average of galaxies radial
velocity is very close to zero, MM2 model (center) shortly after 5 Gyr. In this case is well visible the
increasing of the outer galaxies radial velocity. In SP model (right) the evolution achieved the Hubble
time, and the increasing of $v_r$ is more evident.

\begin{table}[t]
\caption{Left: Physical processes considered in simulations and their evolutionary ages. Right: Values of $H_L$ found from different models from both the best fit of the Hubble diagrams and the average for MM2 and SP models.}
\label{tab:1}
\begin{minipage}{0.5\linewidth}
\vspace{0.4cm}
\begin{center}
\begin{tabular}{|c|c|c|c|}
\hline
Model & $\Lambda$ & ICM & $T_{ev} (Gyr)$ \\
\hline
MM1 & X & & 3.7 \\
MM2 & X & X & 5.3 \\
MM3 &  & X & 3.5 \\
MM4 &  & & 3.5 \\
SP & X & X & 30 \\
\hline
\end{tabular}
\end{center}
\end{minipage}
\hfill
\begin{minipage}{0.5\linewidth}
\vspace{0.4cm}
\begin{center}
\begin{tabular}{|c|c|c|}
\hline
Model & $H_L$ (best fit) & $H_L$ (average) \\
& km/s/Mpc & km/s/Mpc \\
\hline
MM2 & 15.9 $\pm$ 5.3 & 22.0 $\pm$ 1.7\\
SP (5.3 Gyr) & 20.8 $\pm$ 7.1 & 25.0 $\pm$ 4.4 \\
SP (13.5 Gyr) & 41.5 $\pm$ 4.4 & 44.0 $\pm$ 1.3 \\
SP (30 Gyr) & 61.1 $\pm$ 0.7 & 55.4 $\pm$ 0.7 \\
\hline
\end{tabular}
\end{center}
\end{minipage}
\end{table}

\section{Conclusion}

Our results suggest that a kind of LHF is possible only if DE contribution is take into account. The
evaluation of the linear slope in the velocity-distance diagram, allows us to constraint the value of
the local Hubble constant in the range $30 < H_L < 60$ km/s/Mpc.
In the SP simulation we found that the averaged value of $H_L$ increases in time and marks the
evolutionary ages of the cluster, reaching a value of $H_L$ = $55.4 \pm 0.7$ km/s/Mpc at 30 Gyr. From the
best fit of the Hubble diagram we obtain $H_L = 61.1 \pm 0.7$ km/s/Mpc.
Hence, our results suggest that  can be used for determining a rough estimate of the cluster dynamical
age and, at the same time, they can be used for further testing different cosmological scenarios.


\end{document}